\documentclass[conference]{IEEEtran}
\IEEEoverridecommandlockouts
\usepackage{cite}
\usepackage{fancyhdr, amsmath,amssymb,amsfonts,bm}
\usepackage{tipa}
\usepackage{gensymb}
\usepackage{algorithmic}
\usepackage{graphicx}
\usepackage{textcomp}
\usepackage{xcolor}
\usepackage[linesnumbered,ruled]{algorithm2e}
\DeclareMathOperator*{\argmin}{argmin}
\newcommand*{\argminl}{\argmin\limits}
\ifCLASSOPTIONcompsoc
\usepackage[caption=false,font=normalsize,labelfon
t=sf,textfont=sf]{subfig}
\else
\usepackage[caption=false,font=footnotesize]{subfi
g}
\fi
\newcommand{\norm}[1]{\left\lVert #1 \right\rVert}
\def\BibTeX{{\rm B\kern-.05em{\sc i\kern-.025em b}\kern-.08em
    T\kern-.1667em\lower.7ex\hbox{E}\kern-.125emX}}
  
\begin{document}

\title{Cooperative Multi-Modal Localization in Connected and Autonomous Vehicles}

\author{\IEEEauthorblockN{1\textsuperscript{st} Nikos Piperigkos}
\IEEEauthorblockA{\textit{Computer Engineering}\\
\textit{\& Informatics Department}\\
\textit{University of Patras, Greece}\\
piperigkos@ceid.upatras.gr}
\and
\IEEEauthorblockN{2\textsuperscript{nd} Aris S. Lalos}
\IEEEauthorblockA{\textit{Industrial Systems Institute} \\
\textit{Athena Research Center}\\
Patras, Greece \\
lalos@isi.gr}
\and
\IEEEauthorblockN{3\textsuperscript{rd} Kostas Berberidis}
\IEEEauthorblockA{\textit{Computer Engineering}\\
\textit{\& Informatics Department}\\
\textit{University of Patras, Greece}\\
berberid@ceid.upatras.gr}
\and
\IEEEauthorblockN{4\textsuperscript{th} Christos Anagnostopoulos}
\IEEEauthorblockA{\textit{Industrial Systems Institute}\\
\textit{Athena Research Center}\\
Patras, Greece \\
anagnostopoulos@isi.gr}
}

\maketitle

\begin{abstract}
Cooperative Localization is expected to play a crucial role in various applications in the field of Connected and
Autonomous vehicles (CAVs). Future 5G wireless systems are expected to enable cost-effective Vehicle-to-Everything (V2X) systems, allowing
CAVs to share with
the other entities of the network the data they collect and measure. Typical measurement models usually
deployed for this problem, are absolute position from Global
Positioning System (GPS), relative distance and azimuth angle to neighbouring vehicles, extracted from Light Detection and
Ranging (LIDAR) or Radio Detection and Ranging (RADAR) sensors. In this
paper, we provide a  cooperative localization approach that performs multi
modal-fusion between the interconnected vehicles, by representing a fleet of connected cars as an undirected graph, encoding each vehicle position relative to its neighbouring vehicles. This method is based
on: i) the Laplacian Processing, a Graph
Signal Processing tool  that allows to capture intrinsic geometry of the undirected graph of vehicles rather than their absolute position on global coordinate system and ii) the temporal coherence due to motion patterns of the moving vehicles.
\end{abstract}


\section{Introduction}
\label{intro}
Localization  is  one  of the  main  pillars of  Intelligent Transportation Systems (ITS). Although, Global Navigation Satellite System (GNSS), e.g GPS, provide absolute position information, their accuracy is limited and deviations up to 10 m or higher may arise \cite{b8}, \cite{b10}, especially in harsh environments such as urban canyons and tunnels. Since the localization error in autonomous driving should be no greater than 5 m in the worst case,  accurate localization methods must be developed. In recent years, there is a growing interest in Cooperative Localization (CL) as a means to improve GPS accuracy. CL is based on the 5G communication technology V2X, allowing the vehicles of a Vehicular-Ad-hoc-NETwork (VANET) to share information among them, in order to improve the position accuracy. Useful information could be absolute position, relative distance and azimuth angle to neighbouring and connected vehicles.

Multi-modal CL in VANET’s is a research area closely related to CL in Wireless Sensor Networks (WSN). An overview of multi-modal CL in WSN is provided in \cite{b1}. Recent approaches of CL on VANETs, include \cite{b2} where an objective function is formulated by the Maximum Likelihood Estimation criterion and is minimized by employing Distributed Alternating Direction Method of Multipliers. In \cite{b3}, vehicles share absolute position, relative position and motion state measurements and Distributed CL is performed by a Covariance Intersection Filter (CIF), integrating those informations. The VANET of \cite{b4}, fuses absolute position and range measurements, using Extended Kalman Filter (EKF) and CIF, in a Distributed manner. In \cite{b5}, a Distributed CL method in tunnels, that fuses V2X measurements using Particle Filtering, is presented. In \cite{b6}, a Distributed CL method in urban canyons, that fuses absolute position and range measurements using EKF is proposed. In \cite{b7}, a Centralized CL method
based on Generalized Approximate Message Passing
and Kalman Filter (KF) is developed. It exploits navigation measurements from 
Inertial Navigation Unit, GPS, signals of opportunity, ground stations and
 inter-vehicular measurements. In \cite{b8}, a Distributed method is developed, where GPS and vehicles dynamics measurements are fed into a KF and combined with any information available about surrounding features. In \cite{b9}, a Distributed Bayesian approach that fuses GPS and inter-vehicle distance measurements, is employed in order to perform CL. Finally, \cite{b10} provide an overview for cooperative and non-cooperative localization techniques in VANETs.

The previously discussed methods, focus only on the pair-wise measurements of the interconnected vehicles. To the best of our knowledge, the method proposed here is the first one which treats the VANET as a graph and uses graph regulizers to fuse different measurement modalities. Moreover, high-dimensional location trajectories often lie in a low-dimensional subspace and can be recovered more accurately when using exact low-rank matrix recovery approaches methods. We exploit this property of motion patterns using low-rank modelling and optimization tools.

Our main contributions can be summarized as follows:
\begin{itemize}
    \item A novel method for efficient CL is proposed. Our method performs cooperative multi-modal fusion and uses different graph and low-rank modelling tools for exploiting spatial and temporal coherences of moving vehicles.
    \item The new method is implemented in a centralized fashion, assuming the existence of a fusion center.
    \item As shown via extensive experiments, the reduction of Mean Square Localization Error with respect to GPS, can reach 94\%.
\end{itemize}
The rest of the paper is organized as follows:
Section \ref{laplacian} presents the graph regulizer used for exploiting spatial coherences; Section \ref{lrmr} presents the batch sequential solution that exploits the low-dimensional subspace of the high-dimensional location data; Section \ref{simulation} is dedicated to the experimental setup and simulation results while Section \ref{conclusion} concludes our work.

\section{Laplacian based Localization}
\label{laplacian}

The approach that will be derived in this Section, is based on a proper extension of the Laplacian Processing technique \cite{b11}. The motivation was related to estimating vehicles' locations through their differential properties, using LIDAR or RADAR. These differential properties are also derived from certain linear operators defined on the graph that represents the VANET. These two observations motivated us to design graph regulizers that are used for fusing different modalities and exploit spatial coherences in a centralized manner.

\begin{figure}[htbp]
\centerline{\includegraphics[width=0.58\linewidth]{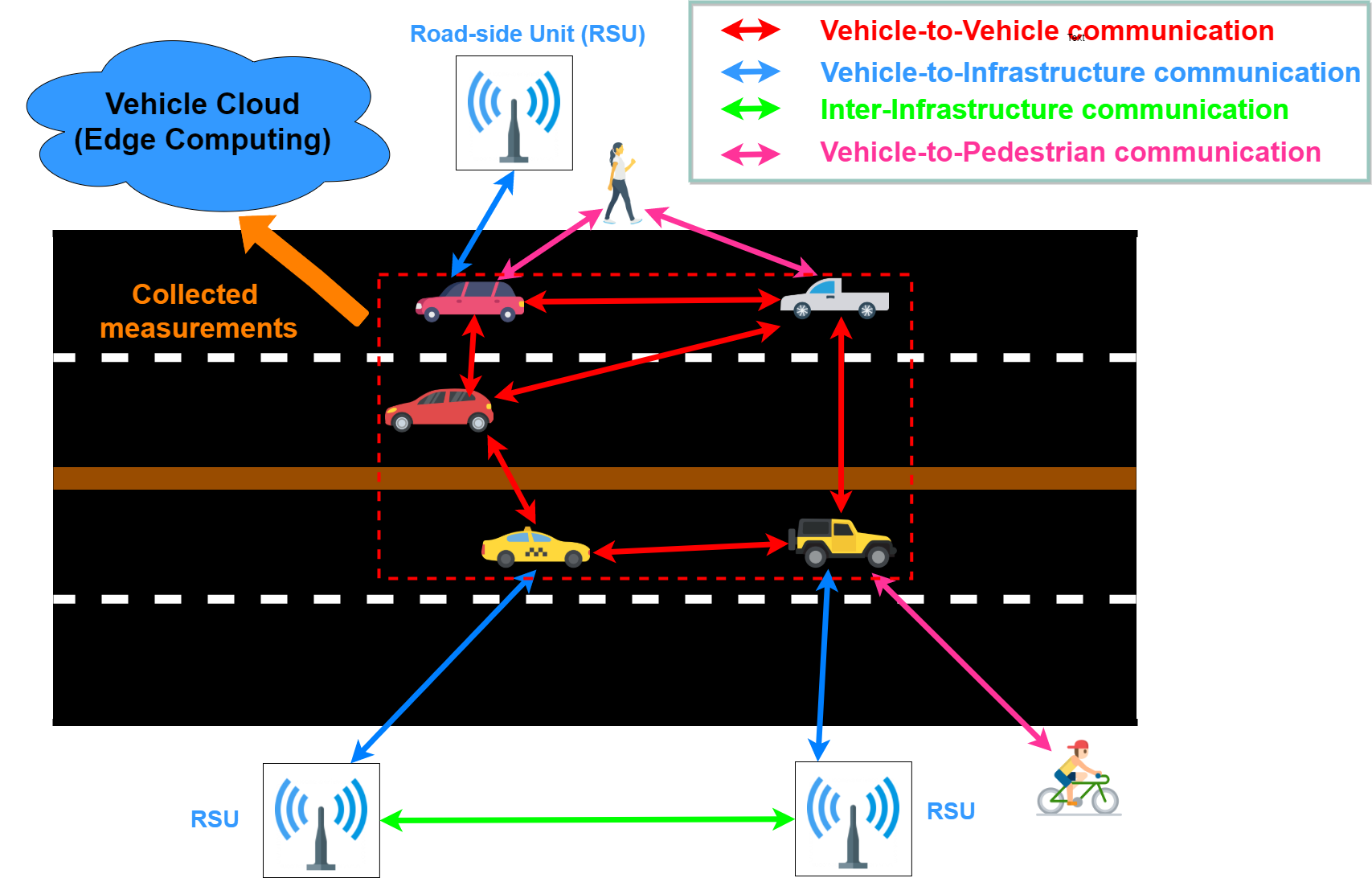}}
\caption{Example of VANET}
\label{fig1}
\end{figure}

\begin{figure}[htbp]
\centering{\includegraphics[width=0.4\linewidth]{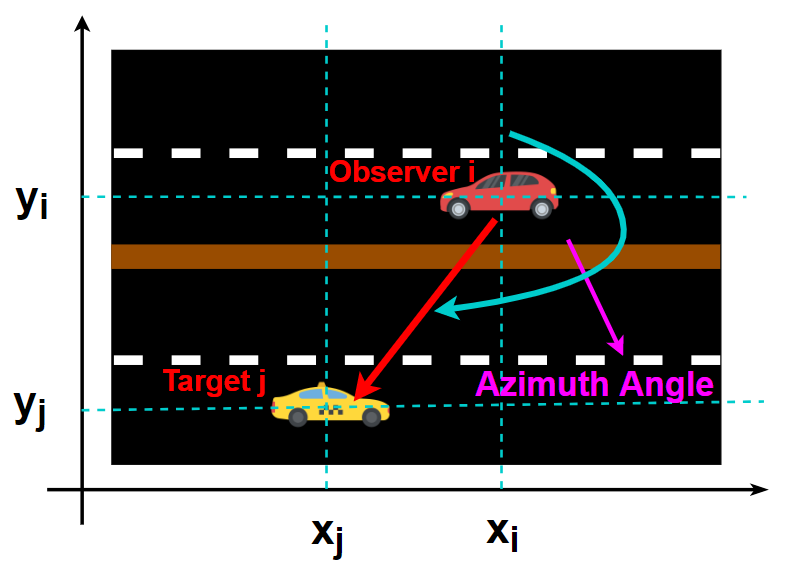}}
\caption{Azimuth Angle measurement}
\label{fig2}
\end{figure}

Consider a 2-D region where N connected vehicles collect measurements while moving. An example of such a VANET, is shown in Fig.~\ref{fig1}.
The location of the $i$-th vehicle at $k$-th time instant is given by $\boldsymbol{x_i^{(k)}} = 
\begin{bmatrix}
x_i^{(k)} & y_i^{(k)}
\end{bmatrix}
$. Each vehicle knows its absolute position from GPS and measures its relative distances and azimuth angles with respect to neighbouring and connected vehicles using LIDAR or RADAR. The true relative distance between connected vehicles $i$ and $j$ is given by: 
     $z_{d,ij}^{(k)} = \norm{{\boldsymbol{x_j^{(k)}} - \boldsymbol{x_i^{(k)}}}}_2$
, where $\norm{\cdot}_2$ is the $l_{2}$ norm. The true azimuth angle (shown in Fig.~\ref{fig2}) between connected vehicles $i$ (observer) and $j$ (target) is given by:
         $z_{az,ij}^{(k)} = \begin{cases}
         \lambda\pi + \arctan{\frac{|x_j^{(k)} - x_i^{(k)}|}{|y_j^{(k)} - y_i^{(k)}|}} , \
         \lambda = 0,1 \\ 
         \lambda\pi + \arctan{\frac{|y_j^{(k)} - y_i^{(k)}|}{|x_j^{(k)} - x_i^{(k)}|}} , \ \lambda = \frac{1}{2}, \frac{3}{2}
         \end{cases}$.
For example, in Fig.~\ref{fig2}, $\lambda = 1$. The acquired measurements are assumed to be described by the following models:
\begin{itemize}
    \item Distance measurement: 
    \begin{align}
    \begin{split}
    \label{eq:1}
        \Tilde{z}_{d,ij}^{(k)} = z_{d,ij}^{(k)} + w_d^{(k)}, \ w_d^{(k)} \sim \mathcal{N}(0,\sigma{_d^2})
    \end{split}
    \end{align}
    \item Azimuth Angle measurement:
    \begin{align}
    \begin{split}
    \label{eq:2}
       \Tilde{z}_{az,ij}^{(k)} = z{_{az,ij}^{(k)}} + w_{az}^{(k)}, \ w_{az}^{(k)} \sim \mathcal{N}(0,\sigma{_{az}^2})
    \end{split}
    \end{align}
    \item Absolute position measurement: 
    \begin{align}
    \begin{split}
    \label{eq:3}
        \boldsymbol{\Tilde{z}_{p,i}^{(k)}} = \boldsymbol{x_i^{(k)}} + \boldsymbol{w_p^{(k)}}, \ \boldsymbol{w_p^{(k)}} \sim \mathcal{N}(0,\boldsymbol{\Sigma{_p}})
    \end{split}
    \end{align} Covariance matrix $\boldsymbol{\Sigma{_p}}$ is a diagonal matrix equal to $diag(\sigma_x^2, \sigma_y^2)$.
\end{itemize}
The noise introduced in the measurements is assumed to be Gaussian, as commonly done in relevant literature \cite{b2}, \cite{b3}, \cite{b4}, \cite{b5}.

Let $G^{(k)} = 
(\mathcal{V}^{(k)} , \mathcal{E}^{(k)})
$ be an undirected graph, which changes over time,  where $\mathcal{V}^{(k)}$ is the set of vertices and $\mathcal{E}^{(k)}$ the set of edges. Each vertex $\boldsymbol{v_i^{(k)}}$ is represented by absolute cartesian coordinates as $\boldsymbol{v_i^{(k)}} = \begin{bmatrix}
x_i^{(k)} & y_i^{(k)}
\end{bmatrix}$. The differential coordinates $\boldsymbol{\delta{_i^{(k)}}}$ for each vertex $\boldsymbol{v_i^{(k)}}$ are defined as:
\begin{align*}
    \begin{split}
        \boldsymbol{\delta{_i^{(k)}}} = [\delta{_i^{(k), (x)}} \ \ \delta{_i^{(k), (y)}}] = \boldsymbol{v_i^{(k)}} - \frac{1}{d_i^{(k)}}{\sum_{j \in N(i)}\boldsymbol{v_j^{(k)}}}
    \end{split}
\end{align*}
, where $d_i^{(k)}$ is the number of neighbours of vertex $\boldsymbol{v_i^{(k)}}$ and $N(i)$ denotes the set of neighbours of $\boldsymbol{v_i^{(k)}}$. Obviously, $\boldsymbol{\delta^{(k), (x)}}, \boldsymbol{\delta^{(k), (y)}} \in \mathbb{R}^{N \times 1}$. We also define the diagonal degree matrix $\boldsymbol{D^{(k)}} \in \mathbb{R}^{N \times N}$ (with $D^{(k)}[i,i] = d_i^{(k)}$ and $D^{(k)}[i,j] = 0$, if $i \neq j$) and adjacency matrix $\boldsymbol{A^{(k)}} \in \mathbb{R}^{N \times N}$ (with $A^{(k)}[i,j]$ = 1, if $(i,j) \in{\mathcal{E}^{(k)}}$ and $A^{(k)}[i,j] = 0$, otherwise). Finally,
the Laplacian Matrix $\boldsymbol{L^{(k)}} \in \mathbb{R}^{N \times N}$ of graph is equal to $\boldsymbol{L^{(k)}} = \boldsymbol{D^{(k)}}-\boldsymbol{A^{(k)}}$. 
Relying on the previously defined $\boldsymbol{\delta^{(k)}}$ coordinates, one can recover the true absolute coordinates of the vertices, represented by the vectors $\boldsymbol{x^{(k)}} \in \mathbb{R}^{N \times 1}$ and $\boldsymbol{y^{(k)}} \in \mathbb{R}^{N \times 1}$, by solving the systems:
\begin{align}
    \begin{split}
    \label{eq:4}
        \boldsymbol{L^{(k)}}\boldsymbol{x^{(k)}} = \boldsymbol{D^{(k)}}\boldsymbol{\delta{^{(k), (x)}}}, \
        \boldsymbol{L^{(k)}}\boldsymbol{y^{(k)}} = \boldsymbol{D^{(k)}}\boldsymbol{\delta{^{(k), (y)}}}
    \end{split}
\end{align}
We can also notice from Fig.~\ref{fig2} that:
\begin{align}
    \label{eq:5}
    \begin{split}
        \delta{_i^{(k), (x)}} = \frac{1}{d_i^{(k)}}{\sum_{j \in N(i)}-\Tilde{z}^{(k)}_{d,ij}\sin{\Tilde{z}^{(k)}_{az,ij}}}
    \end{split}
\end{align}
\begin{align}
    \label{eq:6}
    \begin{split}
        \delta{_i^{(k), (y)}} =  \frac{1}{d_i^{(k)}}{\sum_{j \in N(i)}-\Tilde{z}^{(k)}_{d,ij}\cos{\Tilde{z}^{(k)}_{az,ij}}}
    \end{split}
\end{align}

\begin{figure}[htbp]
\centering{\includegraphics[width=0.38\linewidth]{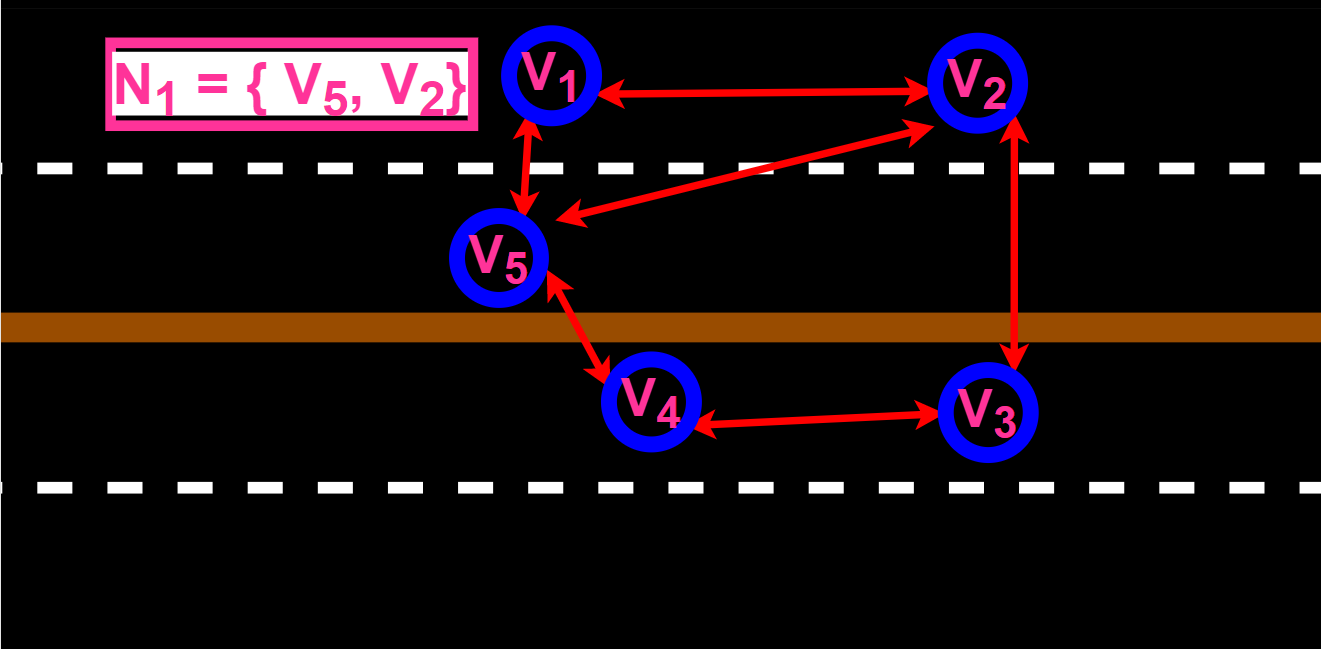}}
\caption{Example of VANET graph}
\label{fig3}
\end{figure}

As such, by considering the vehicles of the network as vertices of a graph and the communication links between the neighbours as its edges, we can create matrices $\boldsymbol{D^{(k)}}, \boldsymbol{A^{(k)}}, \boldsymbol{L^{(k)}}$. The graph that represents the VANET of Fig. \ref{fig1}, is shown in Fig. \ref{fig3}. Moreover, each vehicle utilizing measurement models \eqref{eq:1} and \eqref{eq:2}, can define the $\boldsymbol{\delta_i^{(k)}}$ coordinates and send them to a fusion center, which will try to solve the systems in \eqref{eq:4}.
However, $\boldsymbol{L^{(k)}}$ is a singular matrix, which implies that systems in \eqref{eq:4} are not solvable. Thus, we need to add into the systems some anchor points $\boldsymbol{c_i^{(k)}} =
\begin{bmatrix}
c_i^{(k), (x)} & c_i^{(k), (y)}
\end{bmatrix}
$ with known absolute coordinates. 
Furthermore, if we assume as anchors the vertices $\boldsymbol{v_1^{(k)}}$ and $\boldsymbol{v_5^{(k)}}$, then we define the extended $\boldsymbol{\Tilde{L}^{(k)}}$ which is equal to:
        $\boldsymbol{\Tilde{L}^{(k)}} = \left(\begin{IEEEeqnarraybox*}[][c]{c/c/c} 
        & \boldsymbol{L^{(k)}} 
        \\
        & \boldsymbol{e_1} 
        \\
        & \boldsymbol{e_5} 
       \end{IEEEeqnarraybox*}\right)$
, where $\boldsymbol{e_i} \in \mathbb{R}^{1 \times N}$ is a vector with zeros and $e_i[i] = 1$. If the number of available anchors is $\alpha$, then $\boldsymbol{\Tilde{L}^{(k)}} \in \mathbb{R}^{(N + \alpha) \times N}$. After adding the anchor points, the system in \eqref{eq:4} for $x$-coordinates, can be re-written as:
\begin{align}
    \begin{split}
    \label{eq:7}
        \boldsymbol{\Tilde{L}^{(k)}}\boldsymbol{x^{(k)}} = \boldsymbol{b^{(k),(x)}},    \boldsymbol{b^{(k),(x)}} = \begin{bmatrix} d_1^{(k)}\delta_1{^{(k),(x)}} \ldots    c_{\alpha}^{(k),(x)}
        \end{bmatrix}^T
    \end{split}
\end{align}
Apparently $\boldsymbol{b^{(k),(x)}}\in \mathbb{R}^{(N + \alpha) \times 1}$. 
The required location vector $\boldsymbol{x^{(k)}}$ can be computed in the least-squares sense as follows:
\begin{align}
    \begin{split}
    \label{eq:8}
        \boldsymbol{x^{(k)}} = \left(\boldsymbol{\Tilde{L}^{(k)T}}\boldsymbol{\Tilde{L}^{(k)}}\right)^{-1}\boldsymbol{\Tilde{L}^{(k)T}}\boldsymbol{b^{(k),(x)}} 
        \end{split}
\end{align}


The same approach of \eqref{eq:7} and \eqref{eq:8} is followed for estimating the true $y$-coordinates. In practice, as anchor points the noisy GPS positions of the vehicles of the network (measurement model \eqref{eq:3}) can be used. Furthermore, each $\boldsymbol{\delta{_i^{(k)}}}$ must be multiplied with $d_i^{(k)}$, so as to remain in accordance with \eqref{eq:4}. We  named  our CL approach, \textbf{Graph Regularization for CL} or \textbf{GR-CL}.  The  main steps  of \textbf{GR-CL} are summarized on \textbf{Algorithm \ref{GR-CL}}.

Note that this scheme, can be potentially useful for the self-localization task, where only visually detected road landmarks (e.g. poles, facades) and their relative distance, angle and position to the ego vehicle are exploited. 

\begin{algorithm}
\KwIn{N, T}
\KwOut{$\boldsymbol{x^{(k)}}$, $\boldsymbol{y^{(k)}}$}
\For{$k=1,2,\ldots T$}{
    Movement of vehicles\;
    Create $\boldsymbol{\Tilde{L}^{(k)}}$ from $\boldsymbol{L^{(k)}}$, $\boldsymbol{D^{(k)}}$, $\boldsymbol{A^{(k)}}$\;
  \For {$i=1,2, \ldots N$}{
  $\delta{_i^{(k), (x)}}  = \frac{1}{d_i^{(k)}}{\sum_{j \in N(i)}-\Tilde{z}^{(k)}_{d,ij}\sin{\Tilde{z}^{(k)}_{az,ij}}}$ \;
  $\delta{_i^{(k), (y)}} =  \frac{1}{d_i^{(k)}}{\sum_{j \in N(i)}-\Tilde{z}^{(k)}_{d,ij}\cos{\Tilde{z}^{(k)}_{az,ij}}}$ \;
}
$\boldsymbol{b^{(k),(x)}} = \begin{bmatrix} d_1^{(k)}\delta_1{^{(k),(x)}} \ldots \ c_{\alpha}^{(k),(x)} 
        \end{bmatrix}^T$ \;
$\boldsymbol{b^{(k),(y)}} = \begin{bmatrix} d_1^{(k)}\delta_1{^{(k),(y)}} \ldots \ c_{\alpha}^{(k),(y)} 
        \end{bmatrix}^T$ \;
$\boldsymbol{x^{(k)}} = (\boldsymbol{\Tilde{L}^{(k)T}}\boldsymbol{\Tilde{L}^{(k)}})^{-1}\boldsymbol{\Tilde{L}^{(k)T}}\boldsymbol{b^{(k),(x)}}$ \;  
$\boldsymbol{y^{(k)}} = (\boldsymbol{\Tilde{L}^{(k)T}}\boldsymbol{\Tilde{L}^{(k)}})^{-1}\boldsymbol{\Tilde{L}^{(k)T}}\boldsymbol{b^{(k),(y)}}$ \;
}
\caption{{\bf Graph Regularization for CL or GR-CL} \label{GR-CL}}
\end{algorithm}

\section{Low-rank Matrix Recovery with Laplacian constraints}
\label{lrmr}

%
Kinematic models of vehicles (e.g. bicycle model of \cite{b13}) but also data extracted by open source autonomous driving simulators (e.g. CARLA) exhibit low-rank properties that are attributed to the fact that for various time periods clusters of vehicles move in parallel directions.
Motivated by that, we generated the trajectories of N vehicles for a certain time period and we stored the locations of each time instant to the columns of a matrix. Due to the fact that the columns were correlated and not independent from each other, the matrix proved to be a low-rank. However, if we stored the noisy positions provided by GPS, the matrix will no longer be a low-rank. Based on that facts, we claim that if we were able to recover a low-rank matrix from a batch of continuous but noisy locations, then the true locations would be estimated with high accuracy.

The method of this Section, uses as input the output of the spatial graph regularization of Section \ref{laplacian}. At each $k$-th time instant, \textbf{GR-CL} is implemented and assuming that the fusion center can store the current and the previous $\tau - 1$ $\boldsymbol{\delta_i}$ coordinates and the \textbf{GR-CL} estimations, then matrix $\boldsymbol{B^{(x)}} \in \mathbb{R}^{2N \times \tau}$ can be formulated, where: 
\begin{align}
    \begin{split}
    \label{eq:9}
        \boldsymbol{B^{(k),(x)}} = 
        \begin{pmatrix}
            \boldsymbol{\delta^{(x), (k-\tau)}} & \ldots
            & \boldsymbol{\delta^{(x), (k)}} \\
            \boldsymbol{x^{(k-\tau)}} &  \ldots
            & \boldsymbol{x^{(k)}}
        \end{pmatrix}
    \end{split}
\end{align}
That matrix contains the concatenated vectors $\boldsymbol{b^{(x)}}$ from $(k-\tau)$-th up to $\tau$-th time instant. The \textbf{GR-CL} estimations now act as the anchors. We consider also that the VANET graph does not change while the vehicles are moving, which means that $\boldsymbol{\Tilde{L}}$ remains the same. Thus, a least-squares minimization problem with rank constraints can be formulated: 
\begin{align}
    \begin{split}
    \label{eq:10}
        \argminl_{\boldsymbol{X^{(k)}}}  \norm{\boldsymbol{\Tilde{L}X^{(k)}} - \boldsymbol{B^{(k),(x)}}}_F^2 \ s.t. \ rank(\boldsymbol{X^{(k)}}) <= s 
    \end{split}
\end{align}
, where $\boldsymbol{X^{(k)}} \in \mathbb{R} ^{N \times \tau}$, $\norm{\cdot}_F$ is the Frobenius norm and $s\in[1, min(N,\tau)]$.
Matrix $\boldsymbol{X^{(k)}}$ is in fact the low-rank matrix that correspond to the true $x$-coordinates of the vehicles of the current estimation window of range $\tau$.

In order to solve \eqref{eq:10}, first perform Singular Value Decomposition (SVD) on $\boldsymbol{\Tilde{L}}$, as $\boldsymbol{\Tilde{L}}$ = $\boldsymbol{USV^T}$. Then, create the matrix $\boldsymbol{W^{(k),(x)}}$ = $\boldsymbol{U^TB^{(k),(x)}}$. According to \cite{b12}, matrix $\boldsymbol{X^{(k)}}$ is equal to:
\begin{align}
    \begin{split}
    \label{eq:11}
        \boldsymbol{X^{(k)}}  = \boldsymbol{VS^{-1}}SVT(\boldsymbol{W^{(k),(x)}})_s \
    \end{split}
\end{align}
Operation $SVT(\cdot)_s$ stands for Singular Value Thresholding, which means that only the $s$ largest singular values are used. Apparently, only the last column of $\boldsymbol{X^{(k)}}$ is of interest, since it refers to the locations estimation of the current $k$-th time instant. Solving \eqref{eq:10} requires cubic time complexity, caused by SVD. Therefore a rather small range of estimation window is needed, in order to reduce the computational time. The same approach of \eqref{eq:9}, \eqref{eq:10}, \eqref{eq:11} is followed for estimating the true $y$-coordinates.

As a consequence, a batch sequential method has been developed. It exploits not only the spatially, but also the temporal coherences of moving vehicles. As long as the motion of vehicles is correlated, the temporal coherences are exploited in the sense of determining a low-rank matrix that corresponds to the true location of vehicles. We named this method Graph and Low-Rank Regularization for CL or \textbf{GLRR-CL}, which is summarized on \textbf{Algorithm 
\ref{GLRRCL}}. 

\begin{algorithm}
\KwIn{N, $\boldsymbol{L}$, $\boldsymbol{D}$, $\boldsymbol{A}$, T,  $\tau$, $s$}
\KwOut{$\boldsymbol{X^{(k)}}$, $\boldsymbol{Y^{(k)}}$}
Create $\boldsymbol{\Tilde{L}}$ from $\boldsymbol{L}$, $\boldsymbol{D}$, $\boldsymbol{A}$\;
$\boldsymbol{\Tilde{L}}$ = $\boldsymbol{USV^T}$\;
\For{$k=1,2,\ldots T$}{
  Movement of vehicles\;
  Estimate $\boldsymbol{\delta^{(k)}}$ from \eqref{eq:5}, \eqref{eq:6}\;
  Create $\boldsymbol{b^{(k),(x)}}$, $\boldsymbol{b^{(k),(y)}}$ from \eqref{eq:7}\; 
  Estimate $\boldsymbol{x^{(k)}}$, $\boldsymbol{y^{(k)}}$ from \eqref{eq:8}\;
  Determine $\boldsymbol{B^{(k),(x)}}, \boldsymbol{B^{(k),(y)}}$ using \eqref{eq:9}\;
  \eIf{$k < \tau$}{
    No action is performed\;
   }{
   $\boldsymbol{W^{(k),(x)}}$ = $\boldsymbol{U^TB^{(k),(x)}}$, \
   $\boldsymbol{W^{(k),(y)}}$ = $\boldsymbol{U^TB^{(k),(y)}}$\;
   $\boldsymbol{X^{(k)}} $ = $\boldsymbol{VS^{-1}}SVT(\boldsymbol{W^{(k),(x)}})_s$ \;
   $\boldsymbol{Y^{(k)}} $ = $\boldsymbol{VS^{-1}}SVT(\boldsymbol{W^{(k),(y)}})_s$ \;
   Last columns of $\boldsymbol{X^{(k)}}$ and $\boldsymbol{Y^{(k)}}$ are the final locations estimation at $k$-th time instant\;
  }
 }
\caption{{\bf Graph and Low-Rank Regularization for CL or \textbf{GLRR-CL}} \label{GLRRCL}}

\end{algorithm}

\section{Simulations}
\label{simulation}

\subsection{Experimental Setup}
As explained in Section \ref{lrmr}, the range of the estimation window needs to be relatively small. It is chosen to be $\tau = 10$.
For a reduced computational load, communication links are assumed to exist between the vehicles, only if their distance is lower than 20 m, and the number of connected neighbours is at most 6. We define $\sigma_x$ = 3 m, $\sigma_y$ = 2.5 m, $\sigma_d$ = 1 m, $\sigma_{az} = 4^{\circ}$ and we calculate the Mean Square Localization Error (MSLE) of GPS, \textbf{GR-CL} and \textbf{GLRR-CL}. The Cimulative Distribution Function (CDF) of  MSLE's are presented in Fig.~\ref{fig4}. The conducted experiments were based on: i) Two different types of trajectories, ii) different number of vehicles of the network, iii) different values of $s$. 
It should be noted that the variance of the range measurements error is  usually  much smaller  than  the  variance  of  the  GPS  error, allowing a more accurate estimation of the differential coordinates.

\subsection{Evaluation study}

\begin{figure*}[htbp]
  \centering
  \subfloat[\textbf{GR-CL} with two types of trajectories]{\includegraphics[width=0.25\linewidth]{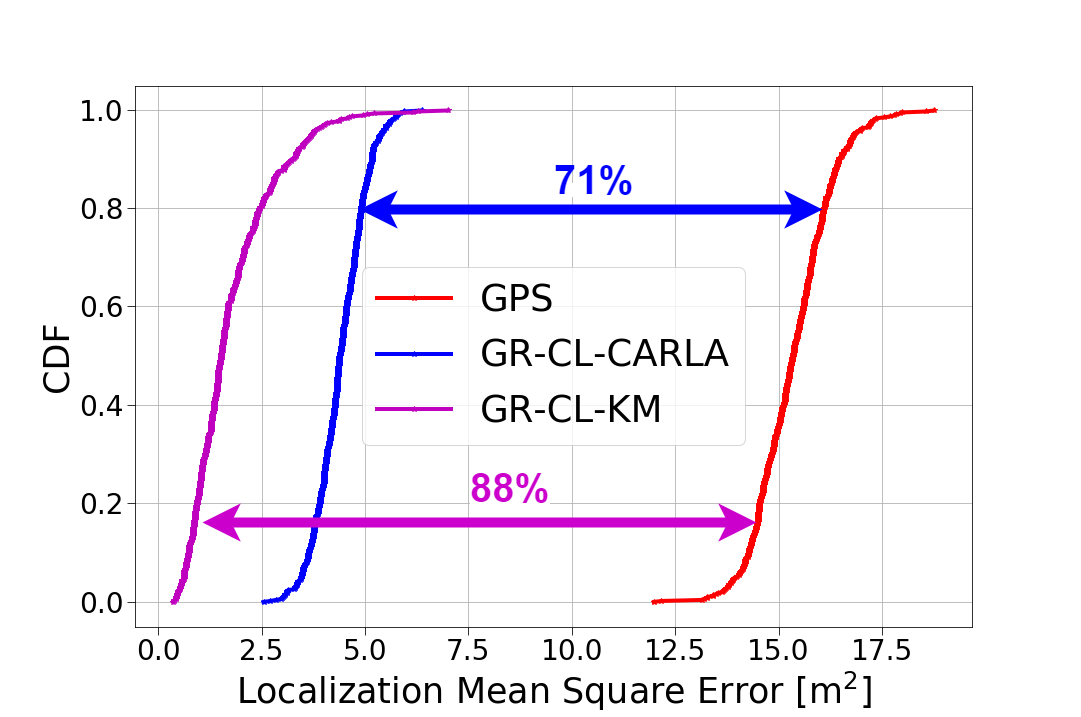}}
  \subfloat[N = 5 , $s = 3$]{\includegraphics[width=0.25\linewidth]{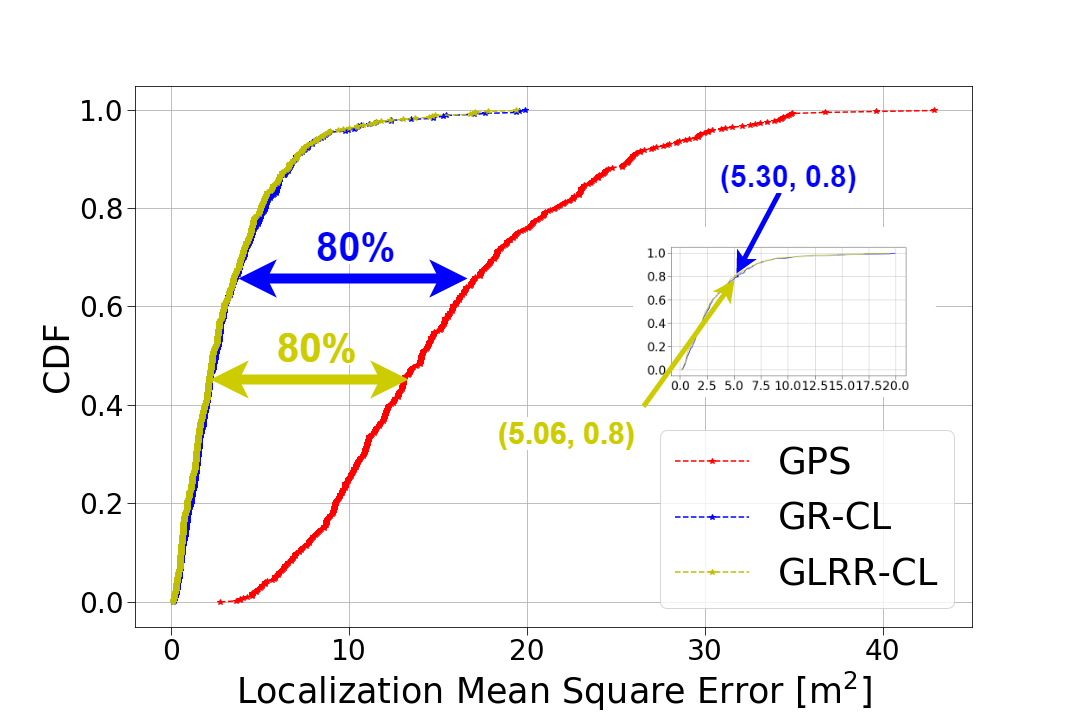}}
  \centering
  \subfloat[N = 25 , $s = 3$]{\includegraphics[width=0.25\linewidth]{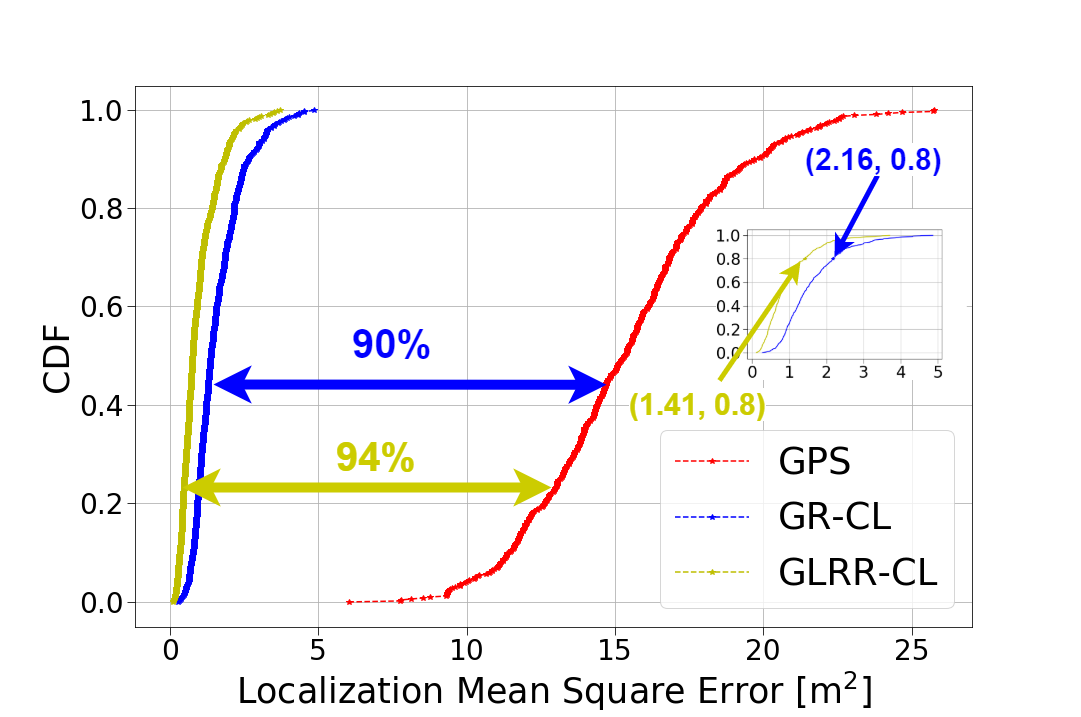}}
  \subfloat[N = 20 , $s = 5$]{\includegraphics[width=0.25\linewidth]{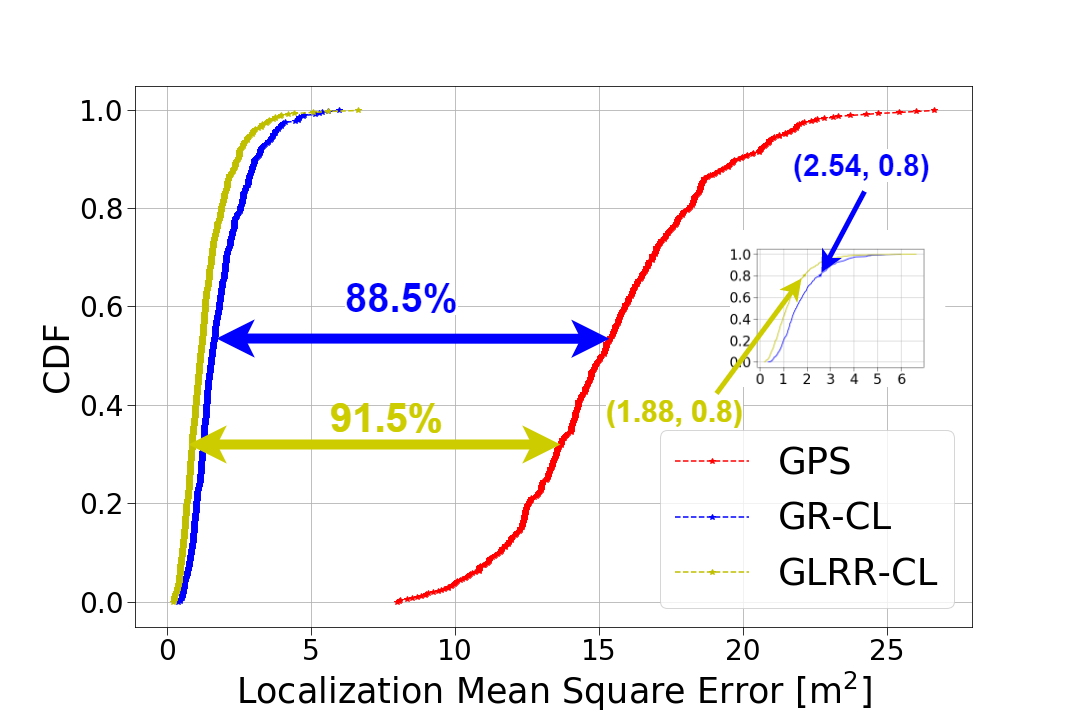}}\\
  \subfloat[N = 20 , $s = 8$]{\includegraphics[width=0.25\linewidth]{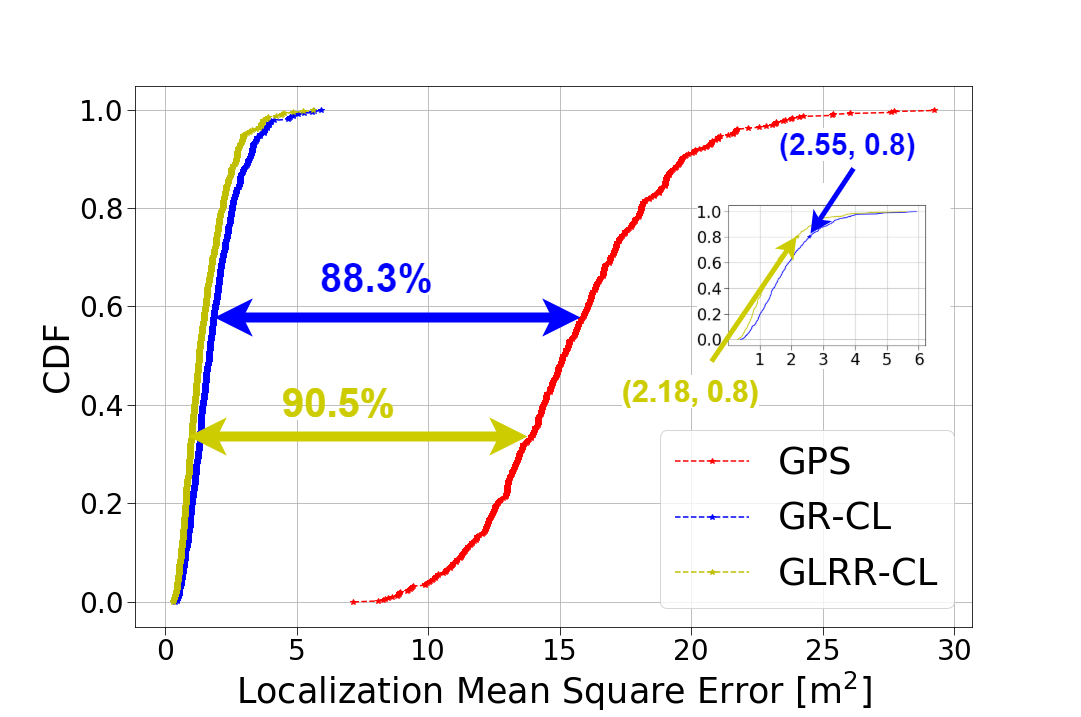}}
  \subfloat[CARLA trajectories]{\includegraphics[width=0.25\linewidth]{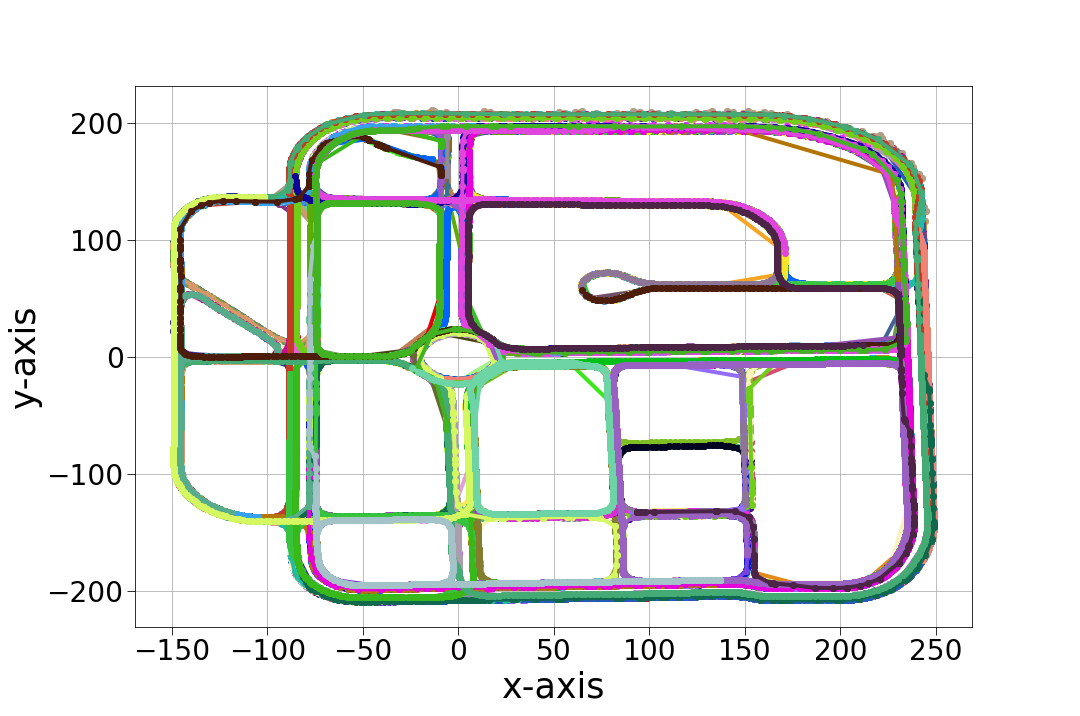}}
  \subfloat[CARLA simulated city]{\includegraphics[width=0.22\linewidth]{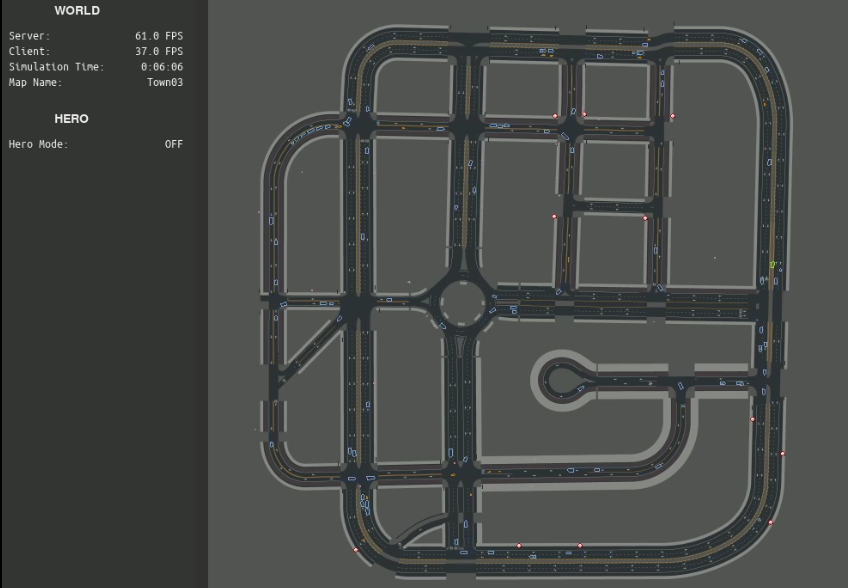}}
  \caption{Mean Square Localization Error and CARLA dataset}
  \label{fig4}
\end{figure*}

Initially, we evaluated the performance of \textbf{GR-CL} on two different types of trajectories for 500 time instances (T = 500). The first one is based on the vehicle kinematic model (KM) of \cite{b13}.
According to that, the current and the previous location is employed in order to generate the trajectories of vehicles:
\begin{align}
    \begin{split}
    \label{eq:12}
        \Acute{x_i}  = x_i + \left( (-s_i/\omega_i)\sin{\theta_i} + (s_i/\omega_i)\sin{(\theta_i + \omega_i\Delta{T})} \right)
    \end{split}
\end{align}
\begin{align}
    \begin{split}
    \label{eq:13}
        \Acute{y_i}  = y_i + \left( (s_i/\omega_i)\cos{\theta_i} + (-s_i/\omega_i)\cos({\theta_i + \omega_i\Delta{T})} \right) 
    \end{split}
\end{align}
\begin{align}
    \begin{split}
    \label{eq:14}
        \Acute{\theta_i}  = \theta_i + \omega_i\Delta{T} 
    \end{split}
\end{align}
, where $\Delta{T}$ is the time step and $\theta_i$, $s_i$, $\omega_i$ are the heading, the speed and the yaw rate of the $i$-th vehicle, respectively. Also, a fixed number of N = 20 vehicles has been used. The second one has been extracted using CARLA. More specifically, we generated the trajectories of N = 300 vehicles shown in Fig.~\ref{fig4}-(f), moving with different directions in a simulated city, shown in Fig.~\ref{fig4}-(g). During the movement and at each time instant, clusters (with modified and not fixed size) of connected vehicles are created. \textbf{GR-CL} is applied on those clusters. The reduction of GPS MSLE in Fig.~\ref{fig4}-(a) is 88\% and 71\% with \textbf{GR-CL} using the KM and CARLA, respectively. \textbf{GR-CL} performs much better when the KM is used. However, the simulated data from CARLA are much closer to real urban conditions. Therefore, it is possible that some vehicles do not have any connected neighbours and thus, do not belong to any cluster. In that case, the error of \textbf{GR-CL} is equal to that of GPS, and \textbf{GR-CL} performs worser.



The different number of vehicles of VANET has also an impact on the performance of the two methods. Assuming the type of trajectory of the KM and that $s$ is optimal and equal to $s= 3$, we generated the trajectories of N = 5 and N = 25 vehicles. The reduction of GPS MSLE in Fig.~\ref{fig4}-(b) is 80\% with \textbf{GLRR-CL} and \textbf{GR-CL}, respectively, while in Fig.~\ref{fig4}-(c) is 94\% and 90\% with \textbf{GLRR-CL} and \textbf{GR-CL}, respectively. It is evident that as the number of vehicles increases, so does the performance of the two methods. \textbf{GR-CL} performs far better at N = 25 than N = 5, because the number of possible connected neighbours of each vehicle grows larger. In that case, \textbf{GR-CL} integrates greater amount of information and performs more efficiently. Regarding the \textbf{GLRR-CL}, we do not notice any benefits when N = 5. In that case the low-rank property doesn't seem to be usefull, since we are trying to recover small matrices with size 5$\times\tau$, where $\tau$ = 10. However, when N = 25, \textbf{GLRR-CL} once again outperforms \textbf{GR-CL}, proving the benefit of exploiting spatio-temporal coherences in the context of a graph regulizer.

In the previous experimental scenario, $s$ was considered known and optimal. Now, we assume that $s$ is not the optimal and more specifically we set $s = 5$ and $s = 8$. Once again, we generated the trajectories of N = 20 vehicles according to the trajectory type of KM. The reduction of GPS MSLE with \textbf{GLRR-CL} in Fig.~\ref{fig4}-(d) is 91.5\%, while in Fig.~\ref{fig4}-(e) is 90.5\%. It is evident that the performance has been degraded when $s$ increases, because in that way we impose to recover a matrix which tends to be a full-rank. \textbf{GLRR-CL} once again outperforms \textbf{GR-CL}.

In all cases \textbf{GLRR-CL} outperforms or experiences the same performance with \textbf{GR-CL}. The key factors of this success is that: i) it exploits  not  only  the  (noisy)  inter-vehicle  measurements  but it  also  integrates  properly  the  connectivity  representation  of VANET  graph, ii) relying on the VANET graph, it benefits from the property that the location data lie in a low-dimensional subspace, allowing the estimation of the low-rank true location matrices. We are planning in the future to apply \textbf{GLRR-CL} also to the simulated datasets of CARLA and to investigate the prospective
of exploiting
road landmarks for the localization task.

\section{Conclusion}
\label{conclusion}

In  this  paper, a two-stage method for CL has been developed. In the first stage, we  modeled  the  vehicles  of a  VANET  as vertices  of an  undirected  graph and  the communication  links between them as its edges. A graph regulizer has been developed, which exploits only spatially coherences, by creating the Laplacian Matrix and defining the $\boldsymbol{\delta}$ coordinates. During the second stage, a graph and low-rank regulizer uses the temporal coherence of vehicles in order to recover a low-rank matrix. An optimization problem with rank and Laplacian constraints has been formulated for the recovery. Different experimental scenarios have proven that the reduction of GPS MSLE with our method can reach 94\%.

\section*{Acknowledgment}
This paper has received funding from the European Union’s H2020 research and innovation programme CPSoSaware under grant agreement No 871738.

\vspace{12pt}

\end{document}